\documentclass[preprint,aps,pra,superscriptaddress,showpacs,floatfix]{revtex4-1}
\usepackage{graphics,epsfig,graphicx}
\usepackage{amsbsy}
\usepackage{bm}
\usepackage{amsfonts}
\usepackage{cancel}
\usepackage{multirow}
\newcommand {\be}{\begin{equation}}
\newcommand {\ee}{\end{equation}}
\newcommand {\bea}{\begin{eqnarray}}
\newcommand {\ea}{\end{eqnarray*}}
\newcommand {\ba}{\begin{eqnarray*}}
\newcommand {\eea}{\end{eqnarray}}

\begin{document}

\title{Universal nature and finite-range corrections in elastic atom-dimer scattering
           below the dimer breakup threshold}

\author{A. Kievsky} 
\affiliation{Istituto Nazionale di Fisica Nucleare, Largo Pontecorvo 3, 56100 Pisa, Italy}
\author{M. Gattobigio}
\affiliation{Universit\'e de Nice-Sophia Antipolis, Institut Non-Lin\'eaire de
Nice,  CNRS, 1361 route des Lucioles, 06560 Valbonne, France }

\begin{abstract}
We investigate universal behavior in elastic atom-dimer scattering below the
dimer breakup threshold calculating the atom-dimer effective-range function
$ak\cot\delta$. Using the He-He system as a reference, we solve the
Schr\"odinger equation for a family of potentials having different values of 
the two-body scattering length $a$ and we compare our results to the universal 
zero-range form deduced by Efimov,
$ak\cot\delta=c_1(ka)+c_2(ka)\cot[s_0\ln(\kappa_*a)+\phi(ka)]$, for selected
values of the three-body parameter $\kappa_*$.
Using the parametrization of the universal functions $c_1,c_2,\phi$ given
in the literature, a good agreement with the universal formula is obtained after
introducing a particular
type of finite-range corrections. Furthermore, we show that the same
parametrization describes a very different system: nucleon-deuteron scattering
below the deuteron breakup threshold. Our analysis confirms the universal
character of the process, and relates the pole energy in the effective-range
function of nucleon-deuteron scattering to the three-body parameter $\kappa_*$.
\end{abstract}

\maketitle

\section{Introduction.}

Scattering of two particles at very low energy shows universal behavior encoded
in the scattering length $a$ and in the effective range $r_s$. In fact,
systems with different interactions sharing the same scattering length and the
same effective range have the same effective range function,
$k\cot\delta=-1/a + r_sk^2/2$, and, accordingly, the same low energy behavior. 
In the limit $a\gg r_0$, where the scattering length is much greater than the 
typical range of the potential $r_0$, not only the scattering process is universal,
but also some bound-state properties. When $a\rightarrow+\infty$ (known as 
unitary limit), the two-particle system has a shallow-bound state with
the bound-state energy $E_2\approx \hbar^2/m a^2$ fixed by the scattering length.
In this limit, the physics is scale invariant.

In the 1970s, V.
Efimov~\cite{efimov:1970_phys.lett.b,efimov:1971_sov.j.nucl.phys.} showed
that the scale invariance is broken in the $s$-wave three-body sector of a
bosonic system. The residual symmetry is the discrete scale invariance (DSI);
namely, the physics is invariant under the rescaling $r\rightarrow
\Lambda^n r$, where the constant is usually written
$\Lambda=\text{e}^{\pi/s_0}$, with $s_0\approx 1.00624$ an universal number
that characterizes a system of three-identical bosons.
One consequence is that at the unitary limit the three-body spectrum consists of
an infinite number of states that accumulate to zero with the ratio between 
two consecutive states being $E_3^{n+1}/E_3^n = \text{e}^{-2\pi/s_0}$. 
For finite scattering length, the binding energies satisfy the
Efimov's equation
\begin{equation}
 E_3^n+\frac{\hbar^2}{ma^2}= e^{-2(n-n_*)\pi/s_0}
\exp{[\Delta(\xi)/s_0]}\frac{\hbar^2\kappa_*^2}{m} \,,
\label{eq:uni3}
\end{equation}
with $\tan\xi=-(mE_3^n/\hbar^2)^{1/2}a$. The function
$\Delta(\xi)$ is universal and a parametrization in the interval 
$[-\pi,-\pi/4]$ is given in Ref.~\cite{braaten:2006_physicsreports}.
The three-body parameter $\kappa_*$ is the wave number 
of the $n=n_*$ state at the unitary limit.

The DSI constrains the form of the observables to be log-periodic functions of
the control parameters. One example is
the atom-dimer scattering length which has the general form
\begin{equation}
a_{AD}/a=d_1 +d_2\tan[s_0\ln(\kappa_*a)+d_3]\,,
\label{eq:a_AD}
\end{equation}
where $d_1,d_2,d_3$ are universal constants 
whose value has been determined in the zero-range
limit~\cite{braaten:2006_physicsreports}.
For collisions below the dimer breakup threshold, DSI
imposes the following universal form for the effective range function
\begin{equation}
ka\cot\delta=c_1(ka) +c_2(ka)\cot[s_0\ln(\kappa_*a)+\phi(ka)]
\label{eq:cotdelta}
\end{equation}
with $\delta$ the atom-dimer phase-shift and $c_1,c_2,\phi$ universal functions
of the dimensionless variable $ka$, where
$k^2=(4/3)E/(\hbar^2/m)$, being $E$ the center of mass energy of the process.
As $k\rightarrow 0$, $ka\cot\delta\rightarrow -a/a_{AD}$ and at $k=0$
the constants $d_1,d_2,d_3$ and $c_1(0),c_2(0),\phi(0)$ are related by simple 
trigonometric relations. A parametrization of the universal constants and 
functions can be found in Ref.~\cite{braaten:2006_physicsreports}.

In this paper we study in detail the universal behavior
of $a_{AD}$ and of the effective range function $ka\cot\delta$.
To this aim, we use the family of atomic $^4$He--$^4$He potentials derived
in Ref.~\cite{gattobigio:2012_phys.rev.a} for several values of $a$, running
from $a\approx 440$ $a_0$ to $a\approx 50$ $a_0$.
The corresponding dimer energies range from $E_2\approx 0.22$ mK to
$E_2\approx21$ mK covering two order of magnitude.
For selected values of $a$ in the mentioned interval we calculate $a_{AD}$ and
the $s$-wave phase-shift in order to construct
the effective range function below the dimer breakup threshold.
As predicted by Eq.(\ref{eq:cotdelta}), when the value of $a$ increases we
observe that $a_{AD}$ changes sign tending to $-\infty$. This behavior produces
a pole in the effective range function. More specifically,
our numerical results are used to analyze the universal form of
Eqs.~(\ref{eq:a_AD}) and (\ref{eq:cotdelta}), and, as the calculations are
done using finite-range interactions, to extract finite-range corrections
by comparing our results to the zero-range theory. Interestingly, for the explored zone
of positive scattering length, the range corrections can be 
taken into account by a shift in the variable $\kappa_*a$. 

This study is of interest for several field of research, ranging from
cold-atoms to nuclear physics and, in particular, in halo nuclei where
a cluster description is justified. In atomic physics, where
the Efimov effect has been observed for the first
time~\cite{kraemer:2006_nature}, discrepancies arise between the theoretical
prediction and the experimental determination of the ratio between $a_*$ and
$a_-$~\cite{zaccanti:2009_natphys,machtey:2012_phys.rev.lett.}, that means
between the scattering lengths at which an Efimov state disappears in the
atom-dimer and in the three-atom continuum, respectively. The solution to this
puzzle is probably hidden in finite-range corrections to the universal formulas
(see Ref~\cite{dyke:2013_arxiv:1302.0281[cond-mat.quant-gas]} and references therein for a recent account of the
problem).
In halo nuclei there has been a lot of interest in the observation of universal aspects
in the scattering of a neutron on a neutron-halo nucleus having
a large scattering length, as for example the $n-^{19}{\rm C}$ system
(see Refs.~\cite{mazumdar:2006_phys.rev.lett.,yamashita:2008_physicslettersb}
and Ref.~\cite{frederico:2012_progressinparticleandnuclearphysics}
for a recent review). In this context we make use of
the universal character of the effective range function to
evaluate a very different system: low energy nucleon-deuteron scattering. It is
well known that the nucleon-deuteron effective range function presents a pole
structure that has been related to the presence of a virtual state. First
observations of this particular behavior have been done in
Refs.~\cite{barton:1969_nuclearphysicsa,whiting:1976_phys.rev.c,girard:1979_phys.rev.c}
whereas in Ref.\cite{chen:1989_phys.rev.c} an explicit calculation of the
effective range function has been done using a semi-realistic nucleon-nucleon
potential. In this last reference the 
calculations allowed to extract the energy of the pole after fitting the effective
range formula with the form suggested by Delves~\cite{delves:1960_phys.rev.}.
In the present work, we show that the pole structure of the effective range
function can be quantitatively related
to the universal form given by Eq.(\ref{eq:cotdelta}) and, using the
parametrization determined in the atomic three-helium system, we apply
that equation to describe nucleon-deuteron scattering as well.
In particular, using the universal function $\phi$, the energy of the pole
can be used to extract the three-body parameter $\kappa_*$.
In this way, the universal behavior imposed by the DSI is analyzed in systems
with natural lengths that differ of several order of magnitude.

\section{The three-boson model.}

We construct the model using the LM2M2
 \cite{aziz:1991_j.chem.phys.}, one of the most used $^4$He-$^4$He potentials,
as the reference interaction, with the mass parameter
$\hbar^2/m=43.281307~\text{($a_0$)}^2\,\text{K}$.
In order to change the value of the scattering length,
we have modified the LM2M2 interaction as
following
\begin{equation}
  V_\lambda(r)=\lambda \cdot V_{\text{LM2M2}}(r)\,\, .
\label{mtbp}
\end{equation}
Examples of this strategy exist in the
literature~\cite{esry:1996_phys.rev.a,barletta:2001_phys.rev.a}.
The unitary limit
is produced for $\lambda\approx 0.9743$ whereas for $\lambda=1$ the values of
the
LM2M2 are recovered: $a=189.41$~$a_0$, $E_2$=-1.303~mK and the
effective range $r_s=13.845$~$a_0$.

Following Ref.~\cite{gattobigio:2012_phys.rev.a}
we define an attractive two-body Gaussian (TBG) potential
\begin{equation}
V(r)=V_0 \,\, {\rm e}^{-r^2/r_0^2}\,,
\label{eq:twobp}
\end{equation}
with range $r_0=10$~$a_0$ and strength $V_0$ fixed to reproduce the values of $a$
given
by $V_\lambda(r)=\lambda \cdot V_{\text{LM2M2}}(r) $.
For example the strength $V_0=-1.2343566$~K corresponds to $\lambda=1$
reproducing the LM2M2 low-energy data,
$E_{2}=-1.303$~mK, $a=189.42$~$a_0$, and $r_s=13.80$~$a_0$.

The use of the TBG potential in the three-atom system produces a
ground-state-binding energy appreciable deeper than the one calculated with
$V_\lambda(r)$.  For example, at $\lambda=1$ the LM2M2 helium-trimer
ground-state-binding energy
is $E_3^0=126.4$~mK whereas the one obtained using the two-body-soft-core
potential in Eq.~(\ref{eq:twobp}) is $151.32$~mK. Hence, we introduce a
repulsive
hypercentral-three-body (H3B) interaction
\begin{equation}
W(\rho_{123})=W_0 \,\, {\rm e}^{-\rho^2_{123}/\rho^2_0}\,,
  \label{eq:hyptbf}
\end{equation}
with the strength $W_0$ tuned to reproduce the trimer energy $E_3^0$ obtained with
$V_\lambda(r)$ for all the explored values of $\lambda$.
Here $\rho^2_{123}=\frac{2}{3}(r^2_{12}+r^2_{23}+r^2_{31})$ is the hyperradius
of three particles and $\rho_0$ gives the range of the three-body force. 
Following Ref.~\cite{gattobigio:2012_phys.rev.a} we use $\rho_0=r_0$.
It should be noticed that the description of the three-boson systems using
a two- plus three-body interaction constructed to reproduce the low energy data
is equivalent, up to finite range corrections, to a description based
on effective field theory (EFT) at leading
order~\cite{bedaque:1999_phys.rev.lett.}.

\begin{figure}
  \begin{center}
  \includegraphics[width=\linewidth]{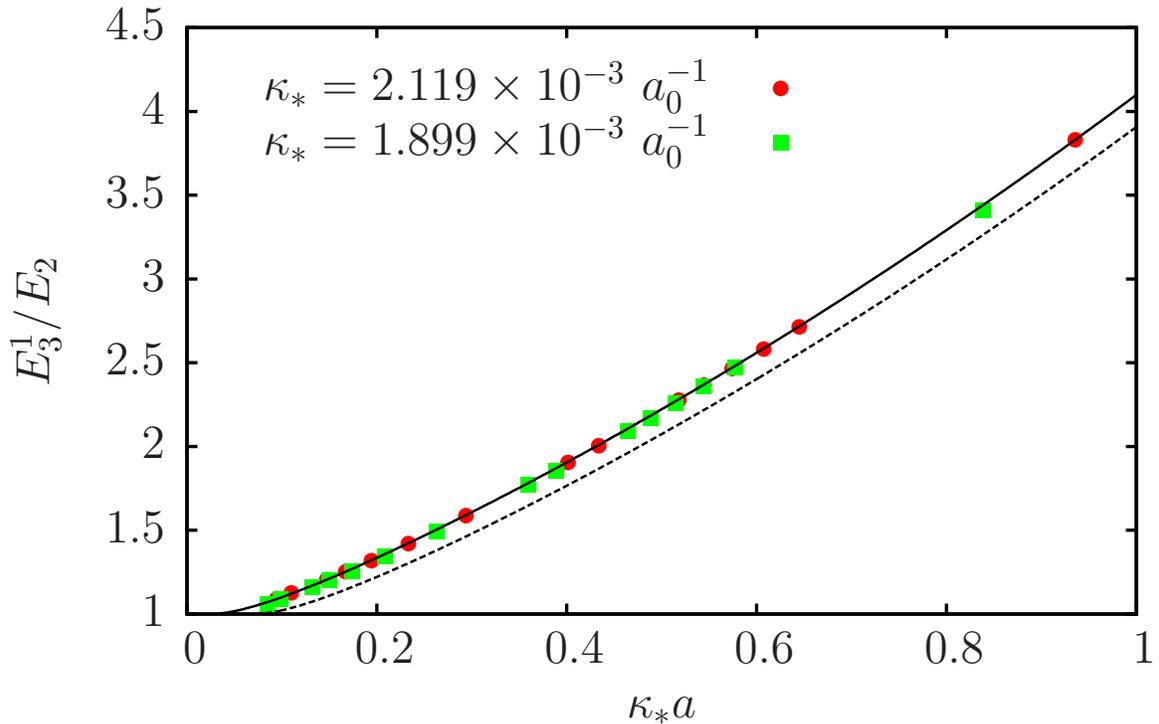}
  \end{center}
  \caption{(Color online) Energy of the first excited state of the trimer as a
  function of $\kappa_*a$. The dashed line is the universal prediction of the
  Efimov law, while the solid line is the translated universal curve.
  The full circles and full squares are the
  calculations using the TBG and TBG+H3B potentials respectively.}
\label{fig:plotnew}
\end{figure}

Varying $\lambda$ from the unitary limit to $\lambda=1.1$ we obtain a
set of values for the ground state binding energy $E_3^0$ and first excited 
state $E_3^1$ using the TBG and TBG+H3B potentials in a broad interval of $a$.
We use the results for $E_3^1$ at the unitary limit (by means of Eq.(\ref{eq:uni3})
with $n_*=1$) 
to determine $\kappa_*=0.002119$~$a_0^{-1}$ and $\kappa_*=0.001899$~$a_0^{-1}$ for the
TBG and TBG+H3B, respectively. In Fig.~\ref{fig:plotnew} we collect
our results for the ratio $E_3^1/E_2$ as a function of $\kappa_*a$ for the
TBG potential (full circles) and of the TBG+H3B potential (full squares). 
We compare the numerical calculations to the predictions of the Efimov's 
binding energy Eq.(\ref{eq:uni3}), given in the figure
by the dashed line.
We observe that the numerical results lie on a curve shifted
with respect to the dashed line. We can interpret the shift as a
consequence of the finite-range character of the numerical results.
Accordingly, we can adapt the Efimov's equation
to treat finite-range interactions. Once we fix $n_*=1$, 
Eq.~({\ref{eq:uni3}}) can be rewritten as follow
\begin{equation}
 E_3^1/E_2=\tan^2\xi\,,\quad
\kappa'_*a=\exp{[-\Delta(\xi)/2s_0]}/\cos\xi \,,
\label{eq:uni31}
\end{equation}
where we have introduced  finite-range correction  by  
$\hbar^2/ma^2 \rightarrow E_2$, in the two-body sector, and 
by a shift $\kappa'_*a = \kappa_*a  + \Gamma$.
Our calculations have been made for two different values of $\kappa_*$,
allowing us to infer that in first approximation $\Gamma\approx \kappa_*r_*$,
with $r_*=21$ $a_0$ $\approx 2 r_0$.
At the unitary limit, the relative weight of this shift becomes negligible,
and Eq.~(\ref{eq:uni31}) tends to Eq.~(\ref{eq:uni3}).

Equation~(\ref{eq:uni31}) can be generalized to states different from the
first excited; however, the shift in this case will also depend on the size
of the state and will not be negligible at the 
unitary limit. 
The corresponding results, obtained by
solving Eq.~(\ref{eq:uni31}) with $\kappa'_*a=\kappa_*a+\kappa_*r_*$,
are shown in Fig.~\ref{fig:plotnew} as a solid line.

\section{atom-dimer scattering.}

To describe atom-dimer scattering below
the dimer threshold we calculate $a_{AD}$ and the $s$-wave atom-dimer phases
$\delta$ using the TBG and TBG+H3B potentials at different energies. We use
the hyperspherical harmonic (HH) method in conjunction with the Kohn
variational principle~\cite{kievsky:1997_nuclearphysicsa}. Applications
of the method to describe a three-helium system with soft-core interactions
as used here can be found in Ref.~\cite{kievsky:2011_few-bodysyst.}.
As an example in Table~\ref{tab:table1} the $a_{AD}$ convergence pattern is
given
for $\lambda=1$ using the TBG potential. The results for $a_{AD}$
are studied in terms of the size of the HH basis given by the grand orbital
quantum
number $K$ and the quantum number $m$ of the hyperradial basis (here taken as
Laguerre functions~\cite{kievsky:1997_nuclearphysicsa}).
When the TBG+H3B potential is considered the
rate of convergence remains the same and the final value is 208 $a_0$.
It should be noticed that as $a$ increases, and correspondingly $|E_2|$
diminishes, is necessary to increase the size of the basis. At the largest
value of $a$ considered, $a\approx 441$ $a_0$, we have used a basis with $m=60$
and $K=800$.

\begin{table}[h]
\caption{ The atom-dimer scattering length $a_{AD}$ (in $a_0$) in terms of the
number
of Laguerre polynomial $m$ and the grand angular quantum number $K$}
\begin{tabular}{c c c c c}
 $m/K$      & 120     & 240      &  360  & 480  \\
\hline
24   & 169.64  & 165.40 & 165.23 & 165.22 \\
28   & 169.59  & 165.32 & 165.10 & 165.09 \\
32   & 169.55  & 165.28 & 165.05 & 165.04 \\
36   & 169.54  & 165.27 & 165.04 & 165.02 \\
40   & 169.53  & 165.26 & 165.03 & 165.02 \\
\hline
\end{tabular}
\label{tab:table1}
\end{table}

\begin{figure}[h]
\bigskip
  \includegraphics[width=\linewidth]{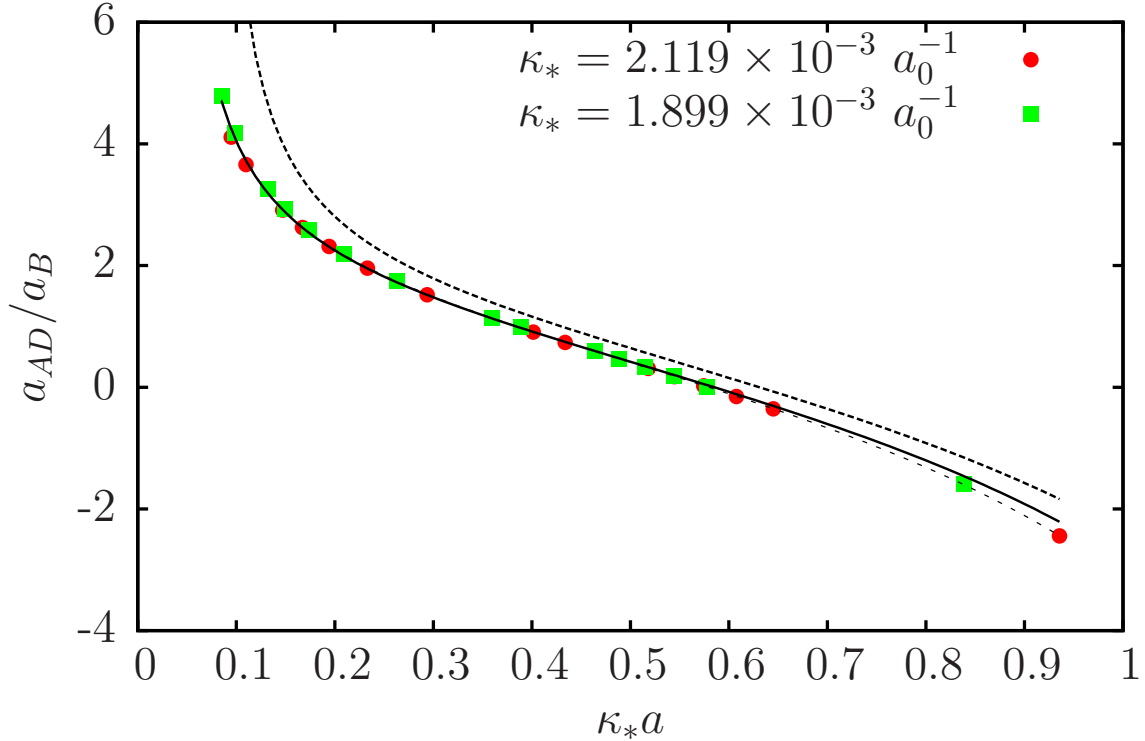}
  \caption{(Color online) Universal plot for $a_{AD}/a_B$ in terms of $\kappa_*a$. 
Open circles
and open squares correspond to TBG and TBG+H3B potentials respectively. 
The dashed line corresponds to Eq.~(\ref{eq:a_AD}), whereas the solid line
corresponds to Eq.~(\ref{eq:a_ADB}). The dotted line 
shows the present parametrization of Eq.~(\ref{eq:a_ADB}).}
\label{fig:a_AD}
\end{figure}

Defining
$E_2=\hbar^2/ma^2_B$, in Fig.~\ref{fig:a_AD} we show the results for the
ratio $a_{AD}/a_B$ in terms of the product $\kappa_* a$. 
It can be observed that the calculated points, 
given as full circles (TBG potential) and full squares (TBG+H3B potential),
lie on a curve shifted with respect to the dashed line representing Eq.(\ref{eq:a_AD})
with the parametrization of Ref.~\cite{braaten:2006_physicsreports}. 
We can interpret again the shift as produced by the finite-range character
of the calculations. Accordingly we can adapt Eq.(\ref{eq:a_AD}) to 
describe finite-range interaction as
\begin{equation}
a_{AD}/a_B=d_1 +d_2\tan[s_0\ln(\kappa'_*a)+d_3]\,.
\label{eq:a_ADB}
\end{equation}
In fact, replacing in the above equation 
$\kappa'_*a=\kappa_*a+\kappa_*r_*$, with the same numerical value of $r_*$ 
as before, the solid line is obtained in Fig.~\ref{fig:a_AD}. 
The values $d_1 = 1.531$ $d_2=-2.141$, $d_3=1.100$
slightly modify the parametrization of Ref.~\cite{braaten:2006_physicsreports}
to better describe the numerical results. This new parametrization is shown
as a dotted line in the figure where we can observe an improvement 
in the description of the results close to the unitary limit.

The shifted formula can be used to determine the ratio $a_*/a_-$,
where $a_-$ is the scattering length at which the three-body states disappear
into the three-atom continuum, and $a_*$ is the scattering length at which the
three-body states disappear into the atom-dimer continuum. For the potential
models used in this work the values of $a_-$ are given in 
Ref.~\cite{gattobigio:2012_phys.rev.a} whereas the values of $a_*$ can be
extracted by equating the argument of the tangent in Eq.(\ref{eq:a_ADB}) to
$-\pi/2$. Using these inputs we obtain $a_*/a_-\approx -0.32$ for both
$\kappa_*=0.002119$~$a_0^{-1}$ and
$\kappa_*=0.001899$~$a_0^{-1}$.
The zero-range universal formulas Eq.~(\ref{eq:uni3}) and Eq.~(\ref{eq:a_AD}) predict $a_*/a_-
= -1.07$, but recent experimental results give lower values for this 
ratio~\cite{zaccanti:2009_natphys,machtey:2012_phys.rev.lett.}. 
The difference is given by finite-range 
effects~\cite{frederico:1999_phys.rev.a,naidon:2011_comptesrendusphysique,ji:2010_epl,dincao:2009_j.phys.b,thgersen:2008_phys.rev.a},
which in our case are encoded in the shift.

\begin{figure}
  \includegraphics[width=\linewidth]{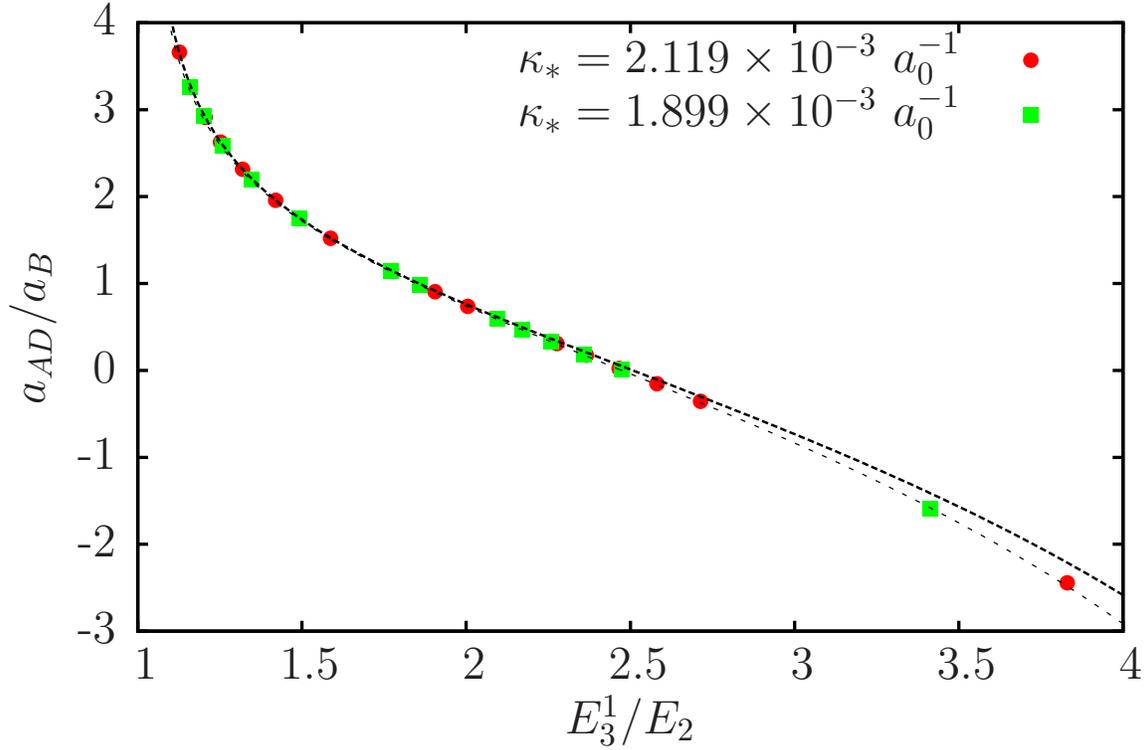}
  \caption{(Color online) The ratio $a_{AD}/a_B$ as a function of $E_3^1/E_2$. Open circles
and open squares correspond to TBG and TBG+H3B potentials respectively.
The dashed line corresponds to Eq.~(\ref{eq:a_AD}) with the parametrization of 
Ref.~\cite{braaten:2006_physicsreports}, whereas the
dotted line shows the present parametrization of Eq.~(\ref{eq:a_ADB}). }
\label{fig:phillips}
\end{figure}

It is interesting to see that the finite-range corrections cancel in the
description of the scattering length as a function of the trimer energy. This is
shown in Fig.~\ref{fig:phillips} in which the present calculations and the
zero-range universal theory,  Eqs.~(\ref{eq:uni3}) and (\ref{eq:a_AD}),
are in close agreement. 

\begin{figure*}
  \includegraphics[width=\linewidth]{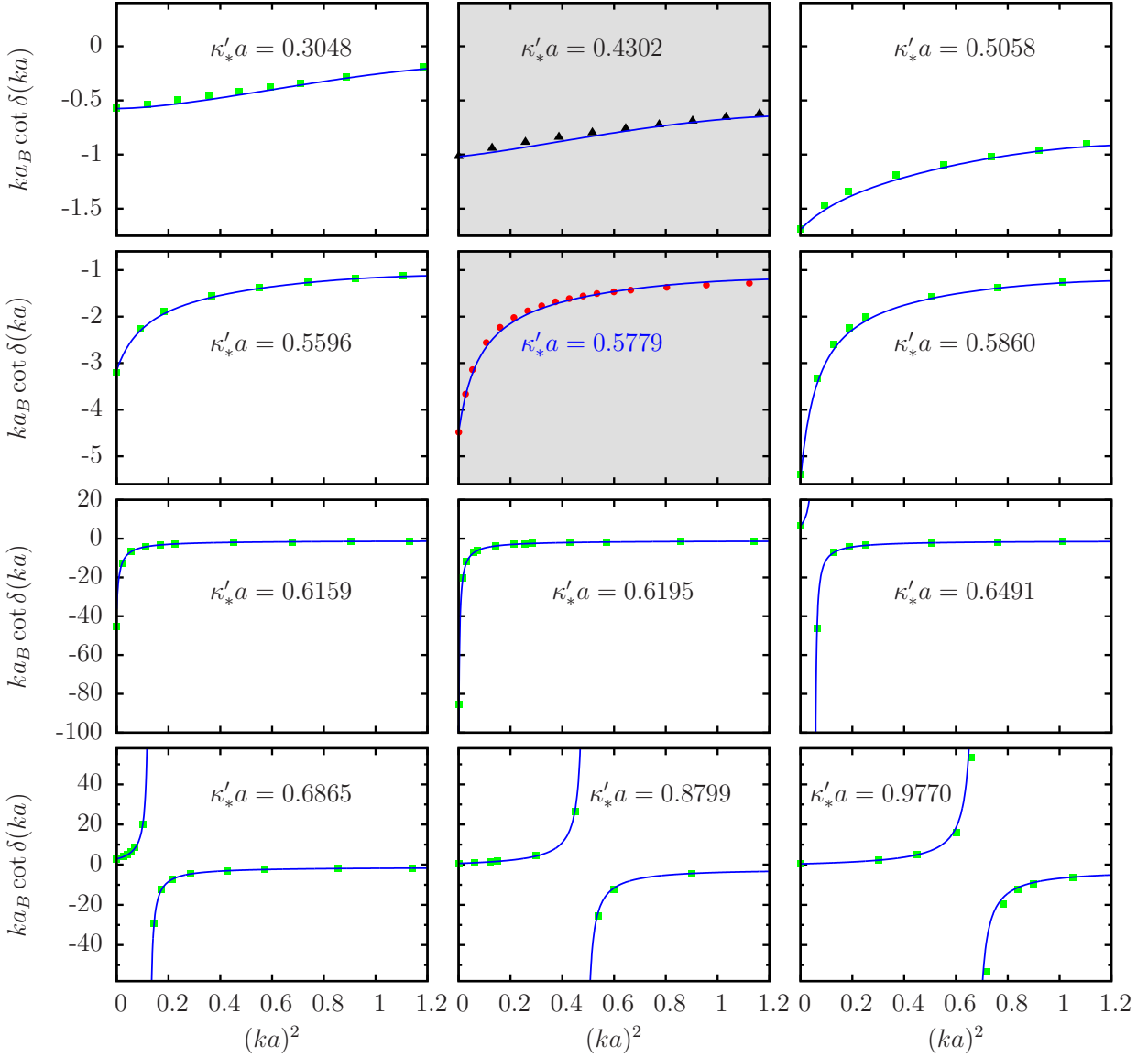}
  \caption{(Color online) The effective range function at different values of $\kappa_*a$  
  as a function of $(ka)^2$. The square points are our calculations; the triangle points
  are the calculations for real  $^4$He-$^4\text{He}_2$ scattering.  The full 
  circles, for $\kappa_*a=0.543055$, correspond to neutron-deuteron scattering
  in the doublet channel.  The solid curves have been calculated using the
  translated-universal formula for the different values of $\kappa_*a$.}
\label{fig:sk}
\end{figure*}

We now present results for atom-dimer scattering at energies below the dimer breakup
threshold for different values of $a$. We adapt Eq.~(\ref{eq:cotdelta}) to
finite-range interactions by considering $ka_B\cot\delta$ as a function of 
the dimensionless center of mass energy $(ka)^2$ and using 
$\kappa'_*a=\kappa_*a+\kappa_*r_*$ in the argument of the logarithm function.
\begin{equation}
    ka_B\cot\delta=c_1(ka) +c_2(ka)\cot[s_0\ln(\kappa_*'a)+\phi(ka)]\,.
  \label{eq:abcotdelta}
\end{equation}
In Fig.~\ref{fig:sk} we show our results (given as full squares) at different
values of $\kappa'_*a$. In the figure 
we can observe very different patterns. For the smallest values of 
$\kappa'_* a$ the behavior is almost linear in all the energy range. 
Starting at values of $\kappa'_*a\approx 0.4$ a curvature appears close to zero
energy, pointing out to an emergent pole structure that becomes evident
at larger values of $\kappa'_*a$.
Specifically, the pole appears 
when $a_{AD}$ changes sign (see Fig.~\ref{fig:a_AD}) or, as given
in Eq.(\ref{eq:abcotdelta}), when the argument of the cotangent
function becomes zero (or $n\pi$). 
The shadow plot in the first row of Fig.~\ref{fig:sk} corresponds to the case
$\lambda=1$ and describe $^4$He-$^4\text{He}_2$ scattering (full triangles). 
The shadow plot in the second row corresponds to nucleon-deuteron scattering as 
discussed below.
The solid curves are obtained using the finite-range-adapted Eq.(\ref{eq:abcotdelta}) with the
parametrization of Ref.~\cite{braaten:2006_physicsreports}. We can
observe a noticeable agreement along the whole range of values.

\section{nucleon-deuteron scattering.}

The universal effective range function has been
determined using the TBG and the TBG+H3B potentials describing an atomic
three-helium system. The universal character of the function allow us to apply
it
to describe a very different system: nucleon-deuteron ($n-d$) scattering.
Many efforts have been given in the past to understand the peculiar
form of the $n-d$ effective range function a low energies (see
Refs.~\cite{barton:1969_nuclearphysicsa,whiting:1976_phys.rev.c,girard:1979_phys.rev.c}).
However, Efimov showed that
this process can be described with the universal formula of
Eq.(\ref{eq:cotdelta})~\cite{efimov:1979_sov.j.nucl.phys.}. Accordingly we would like to
apply Eq.(\ref{eq:abcotdelta}) to perform a quantitatively description of
$n-d$ scattering at low energies.
To this aim we use the results of Ref.~\cite{chen:1989_phys.rev.c} in which
$n-d$ scattering has been described using a spin dependent
central potential. In that reference they obtained $a_{nd}=0.71$ fm for the
$n-d$
scattering length, and the effective range function has been parametrized as
\begin{equation}
k\cot\delta=\frac{-1/a_{nd}+r_sk^2/2}{1+E_{c.m.}/E_p}
\end{equation}
with $E_p=-160$~keV and $r_s\approx-127$~fm. It should be noticed that this particular
parametrization of the effective range function can be simple related to 
Eq.(\ref{eq:abcotdelta}) in the low energy limit. Accordingly,
from the values of $E_p$ and $a_{nd}$ it is possible to determine
the corresponding values of $a$ and $\kappa'_*$ by using the universal function
$\phi$ and Eq.~(\ref{eq:a_ADB}). We obtain $a=4.075$~fm and
$\kappa'_*a=0.5779$. The pole appears in the negative
region similar to what happens in Fig.~\ref{fig:sk} at intermediate
$\kappa'_* a$ values. The shadow panel of the second row in
Fig.~\ref{fig:sk} shows a comparison of
the universal function (solid line) to the $n-d$
scattering results of Ref.~\cite{chen:1989_phys.rev.c} (full circles).
We observe a noticeable agreement.
It should be noticed that the spin dependent potential used
in Ref~\cite{chen:1989_phys.rev.c} reproduces the singlet (${}^1a_{np}\approx -20$ fm)
and triplet (${}^3a_{np}\approx 5$ fm) $n-p$ scattering lengths. Accordingly,
the three-nucleon systems has a symmetric plus a mixed symmetry component.
The value of $a$ extracted from the universal function, which is close to
${}^3a_{np}$, can be considered an effective value in an equivalent
three-boson system with the given $a_{nd}$ and the corresponding
effective range function.
A more deeper analysis extending the model to spin dependent interactions
is at present underway.

This study could be of interest in light neutron halo systems in which a
low energy neutron can impact on a loosely bound n-core system. For example,
specific applications of Eq.(\ref{eq:cotdelta}) recently appears in 
low energy $n-^{19}{\rm C}$ scattering~\cite{yamashita:2008_physicslettersb}.
We expect that the use of Eq.(\ref{eq:cotdelta}) will be useful in
further studies of such systems.

\section{conclusions}

In this work we have analyzed the low energy behavior of 
a three-boson system in which the interaction between two bosons
produces a large scattering length. Following
Ref.~\cite{gattobigio:2012_phys.rev.a},
we have used the three-helium system
as a reference system and, in the spirit of EFT at LO, we have
constructed an attractive two-body interaction plus a repulsive three-body
interaction devised to reproduce the two-body scattering length and
three-body binding energy of the LM2M2 interaction.
From our numerical results we have shown that, when finite range interactions
are used, the zero-range universal formula has to be adapted introducing a
shifted three-body parameter
$\kappa'_*a = \kappa_*a+\kappa_*r_*$. This is a particular type of range
corrections and the numerical value of $r_*\approx 2r_0$, in the case of the
helium system, has been extracted from our calculations. We have proposed
and solved Eq.(\ref{eq:uni31}) to describe the spectrum of a three-boson
system.
This equation can be considered an extension of Eq.(\ref{eq:uni3}) for finite
range interactions; they are characterized by two parameters, the effective
range $r_s$ (in the relation between $E_2$ and $a$) and the shift $r_*$, 
that are somehow connected. Their relation is at present subject of investigation.

In the case of atom-dimer scattering we have proposed Eqs.(\ref{eq:a_ADB}) and
(\ref{eq:abcotdelta}) introducing the shifted parameter $\kappa'_*$. In order
to describe our numerical results we have used the parametrization of
Ref.~\cite{braaten:2006_physicsreports}, which, without the inclusion of the
shift, can not quantitatively describe the very complicated structure of the
effective range function.
Interestingly, the value of $r_*$ necessary to describe the results is the same
for $E_3^1$, for $a_{AD}$ and for the effective range function.
This type of correction can be
compared to the range corrections obtained in very recent works
\cite{platter:2009_phys.rev.a,srensen:2012_phys.rev.a}.

As a second application we have used Eq.(\ref{eq:abcotdelta})
to describe low energy $n-d$ scattering. The values of $\kappa'_*$ and $a$
entering in the equation have been determined from the pole energy $E_p$ and
the doublet scattering length $a_{nd}$ given
in Ref.~\cite{chen:1989_phys.rev.c}. A quantitative agreement between
a direct application of Eq.(\ref{eq:abcotdelta}) and the calculations on the
$n-d$ system has been found. This analysis connects
the universal behavior of atomic systems having large two-body scattering length
to nuclear systems. Work in progress includes the extension of the present
analysis to energies above the dimer breakup, in particular the
description of the recombination rate at threshold, and the extension to include
spin and isospin degrees of freedom in order to consider nuclear systems.

\bibliography{biblio}

\end{document}